\documentclass[lettersize,journal]{IEEEtran}

\usepackage{microtype}
\usepackage{graphicx}
\usepackage{subfigure}
\usepackage{booktabs} 
\usepackage{tikz}
\usetikzlibrary{arrows.meta,positioning,calc,decorations.pathmorphing,fit}

\usepackage{hyperref,comment}

\usepackage{amsmath}
\usepackage{amssymb}
\usepackage{mathtools}
\usepackage{amsthm}

\usepackage[capitalize,noabbrev]{cleveref}

\theoremstyle{plain}
\newtheorem{theorem}{Theorem}
\newtheorem{proposition}[theorem]{Proposition}
\newtheorem{lemma}{Lemma}

\theoremstyle{definition}
\newtheorem{definition}{Definition}

\theoremstyle{remark}

\newcommand{\Heven}{\mathcal{H}_{\text{even}}}
\newcommand{\Hodd}{\mathcal{H}_{\text{odd}}}

\newtheorem{principle}{Principle}

\title{The Geometry of Certainty: Recursive Topological Condensation and the Limits of Inference}

\author{%
  Xin Li,\textsuperscript{1}\thanks{This work was partially supported by NSF IIS-2401748 and BCS-2401398. The author has used Gemini 3.0 and ChatGPT 5.1 models to assist in the development of theoretical ideas and visual illustrations presented in this paper.} \\
  Department of Computer Science\\
  University at Albany\\
  \texttt{xli48@albany.edu}
}
\begin{document}

\maketitle

\begin{abstract}
Computation fundamentally separates time from space: nondeterministic search is exponential in time but polynomially simulable in space (Savitch’s Theorem). We propose that the brain physically instantiates a biological variant (inverse) of Savitch's theorem through Memory-Amortized Inference (MAI), creating a geometry of certainty from the chaos of exploration. We formalize the cortical algorithm as a recursive quotient topology operation of flow into scaffold:$H_{\text{odd}}^{(k)} \xrightarrow{\text{Condense}} H_{\text{even}}^{(k+1)}$, where a stable, high-frequency cycle ($\beta_1$) at level $k$ is collapsed into a static atomic unit ($\beta_0$) at level $k+1$. Through Topological Trinity (Search $\to$ Closure $\to$ Condensation), the system amortizes the thermodynamic cost of inference. By reducing complex homological loops into zero-dimensional defects (memory granules), the cortex converts high-entropy parallel search into low-entropy serial navigation. This mechanism builds a ``Tower of Scaffolds'' that achieves structural parity with the environment, allowing linear cortical growth to yield exponential representational reach. However, the efficiency imposes a strict limit: the same metric contraction that enables \emph{generalization} (valid manifold folding) inevitably risks \emph{hallucination} (homological collapse). We conclude that intelligence is the art of navigating such a fundamental trade-off, where the ``Geometry of Certainty'' is defined by the precise threshold between necessary abstraction and topological error.
\end{abstract}

\section{Introduction}
\label{sec:0}

\noindent\textbf{The Exponential Barrier.}
A central puzzle in cognitive science and theoretical machine learning is how biological intelligence achieves \emph{exponential} problem-solving abilities using only \emph{polynomial} increases in physical substrate. Across mammals, neocortical surface area grows roughly linearly \cite{buzsaki2006rhythms}, yet behavioral sophistication, manifested as the depth of counterfactual reasoning and the height of abstraction, grows superlinearly. How does a resource-limited biological system, constrained by the slow transmission speeds of wetware, acquire the functional equivalent of exponential search?

We propose that the answer lies in the \emph{geometry of certainty}: the brain's ability to transmute the high-entropy time cost of exploration into low-entropy spatial structures. By recursively condensing dynamic flows into static scaffolds, the cortex creates a \emph{Tower of Scaffolds} that mirrors the causal depth of the environment \cite{hawkins2021thousand}, allowing it to \emph{navigate} complexity rather than compute it.
High-level cognition requires navigating a branching tree of counterfactual futures: \emph{If I take this action, what else becomes possible?} Computationally, such tasks correspond to nondeterministic search in complexity classes such as $\mathrm{NPSPACE}$, where the number of hypothetical branches grows exponentially with depth. Minsky identified such a combinatorial explosion as the ``Search Problem'' \cite{minsky1967computation}, arguing that no physically bounded agent can explore the solution space via brute force. The fundamental constraint of intelligence is thermodynamic: checking every path costs too much energy.

\noindent\textbf{The Cortical Strategy: Recursive Condensation.}
We propose that the brain circumvents the exponential barrier via the \emph{Topological Trinity Transformation} (Search $\rightarrow$ Closure $\rightarrow$ Condensation). Under the framework of \emph{Memory-Amortized Inference} (MAI), the cortex does not merely search; it permanently deforms the solution space. We formalize the deformation as a recursive quotient topology operation \cite{gallier2022homology}:
$q: \mathcal{M}^{(k)} \longrightarrow \mathcal{M}^{(k)} / \sim_\gamma$,
where the equivalence relation $\sim_\gamma$ identifies all states along a stable inference path $\gamma$ as a single point. Physically, this represents a metric contraction driving the geodesic distance along the cycle to zero, creating a topological ``wormhole'' that collapses the temporal duration of the thought into a structural instant. The quotient map instantiates the homological transition:
$\mathcal{H}_{\text{odd}}^{(k)} \xrightarrow{\text{Condense}} \mathcal{H}_{\text{even}}^{(k+1)}$,
where a high-frequency cycle ($\beta_1$) at level $k$ is transmuted into a static atomic unit ($\beta_0$) at level $k+1$. Topological condensation thus transforms the time complexity of processing into the space complexity of memory \cite{sipser1996introduction}, converting nondeterministic branching into deterministic navigation.
Such a recursive mechanism explains the linear-to-exponential scaling of the neocortex. By stacking these quotient manifolds, the brain adheres to the \emph{Parity Principle}: the internal hierarchy of condensed nodes ($\beta_0$) constructs a structural isomer of the external world’s causal hierarchy. Consequently, each incremental increase in cortical area does not just add more storage; it adds a new layer of \emph{meta-scaffolding} \cite{bennett2023brief}. The meta-scaffolding strategy allows the organism to reason about ``concepts'' with the same low metabolic cost as reasoning about ``pixels'', effectively acting as an evolutionary cheat code that yields algorithmic power unattainable by unstructured parallel hardware.

\noindent\textbf{The Geometry of Certainty and its Limits.}
Topologically, we posit that the inference process is governed by the boundary-of-a-boundary identity $\partial^2=0$ \cite{wheeler1990information}. Intelligence arises from enforcing \emph{Topological Closure}: ensuring that inference paths form valid cycles (homology classes) before they are condensed. Although geometric rigor creates certainty from probabilistic noise, the inference efficiency comes with a strict limit. The same metric contraction that pulls related concepts together to enable \emph{generalization} (valid manifold folding) \cite{geman1992neural,bengio2013representation} risks \emph{hallucination} (homological collapse) \cite{ji2023survey,beck2003cognitive}. When the manifold is over-condensed, distinct concepts merge, creating certain but erroneous paths. Therefore, the limit of efficient inference is defined by the topological resolution of the condensed space.
This paper develops the theoretical and computational foundations of the Geometry-of-Certainty principle. Our contributions are:
\begin{enumerate}
    \item We formalize the topological trinity as the universal condensation algorithm for converting high-entropy search into low-entropy navigation;
 \item We prove a proposition about recursive condensation, demonstrating how the brain transmutes homological flow ($\beta_1$) into architectural scaffolds ($\beta_0$);
 \item We demonstrate that MAI acts as an inverse to a biological realization of Savitch’s Theorem ($\mathrm{NPSPACE} \subseteq \mathrm{DSPACE}$) \cite{savitch1970relationships}, replacing exponential time with polynomial structural reuse;
 \item We interpret the cortical column as the physical operator of the trinity transformation, vindicating Mountcastle’s universality hypothesis \cite{mountcastle1957modality};
 \item We unify generalization and hallucination as the same topological phenomenon, metric contraction, differentiated only by ground-truth validity.
\end{enumerate}

\section{Foundation: Topological Condensation}
\label{sec:theoretical_foundations}

To understand why the brain adopts a recursive topological architecture, we must first define the thermodynamic problem it solves. Intelligence is not merely the processing of information; it is the physical act of navigating a state space under strict energy constraints. In this section, we formalize cognition as pathfinding on a high-dimensional latent manifold $\mathcal{M}$ and propose a topological amortization strategy based on the homological parity principle.

\subsection{Ergodicity vs. Navigation}
Classical approaches to search, typified by Reinforcement Learning (RL) \cite{sutton2018reinforcement} or Monte Carlo methods \cite{liu2001monte}, are fundamentally ergodic. They rely on the Law of Large Numbers, assuming that if an agent samples the state space $\Omega$ long enough, the time average of its experience will converge to the spatial average.\begin{equation}\lim_{T \to \infty} \frac{1}{T} \int_0^T f(x_t) dt \approx \int_{\Omega} f(x) p(x) dx\end{equation}While mathematically robust, ergodicity is biologically ruinous. It implies that to know a domain, the agent must visit every state or at least sample the volume densely. For a high-dimensional cognitive manifold where $|\Omega|$ scales exponentially with feature count, the time complexity (the convergence time $T$) becomes infinite for all practical purposes. The exponential barrier has been known as the curse of dimensionality in dynamic programming \cite{bellman1966dynamic} or Minsky's search problem in AI \cite{minsky1967computation}. 

We posit that biological systems are non-ergodic. They do not seek to sample the space; they seek to \emph{navigate} it, based on the phylogenetic continuity hypothesis of navigation and memory \cite{buzsaki2013memory}. We define the cognitive task not as estimating a distribution, but as finding a specific geodesic path $\gamma$ between an initial state $s_{start}$ and a goal state $s_{end}$ on a Riemannian manifold $(\mathcal{M}, g)$ - i.e., $\gamma: [0, 1] \to \mathcal{M}$. The goal of the brain is to minimize the action (energy cost) of this path. In an undeformed (unlearned) manifold, the metric tensor $g$ is uniform (flat), and the path is long and uncertain. Learning, in a geometric framework, is the warping of the manifold itself, altering $g$ to shorten the geodesic distance between relevant states.

\subsection{Topological Condensation for Amortization}

We define the cost of certainty as the metabolic action required to compute the inference path $\gamma$. We distinguish between the high-energy thermodynamic state of search (i.e., slow thinking or system 2 \cite{kahneman2011thinking}) and the low-energy ground state of memory (i.e., fast thinking or system 1 \cite{kahneman2011thinking}), connected by a specific metric deformation.

\begin{definition}[The Search Action]
When the system encounters a novel stimulus, the manifold metric $g$ is effectively flat or undefined relative to the goal. The system must perform an active search, modeled as a path integral over the latent landscape. We define the search action $\mathcal{A}_{\text{search}}$ as:
\begin{equation}
\mathcal{A}_{\text{search}}(\gamma)
= \int_{0}^{1} \Big(
\underbrace{\sqrt{g_{ij}\,\dot{\gamma}^i \dot{\gamma}^j}}_{\text{Geodesic Work}}
\;+\;
\underbrace{T \cdot \mathcal{S}(\gamma(t))}_{\text{Entropic Heat}}
\Big)\, dt
\end{equation}
\end{definition}
where the first term is the metric distance (inference depth), and the second term represents the thermodynamic heat generated by information processing (the time cost), where $T$ is the ``cognitive temperature'' (noise/exploration rate) and $\mathcal{S}$ is the local entropy. For non-trivial problems, $\mathcal{A}_{\text{search}}$ scales super-linearly with problem depth (e.g., the maginal number of seven \cite{miller1956magical} in working memory).

\begin{definition}[The Condensation Operator]
We define the condensation operator $\Gamma$ as a map that acts on the manifold's metric tensor $g$ along a validated cycle $\gamma$. If a path $\gamma$ satisfies the closure condition ($\partial \gamma = 0$), $\Phi$ induces a local metric contraction:\begin{equation}\Gamma: g_{ij}(x) \mapsto e^{-\lambda t} g_{ij}(x) \quad \forall x \in \text{Im}(\gamma)\end{equation}where $\lambda$ is the plasticity rate (learning rate) and $t$ is the number of successful traversals. As $t \to \infty$, the metric distance between any two points on the cycle approaches zero.
\end{definition}


\begin{lemma}
\textit{Under the action of the condensation operator $\Phi$, the search action $\mathcal{A}(\gamma)$ converges to a constant lower bound $\epsilon$, effectively decoupling inference cost from problem complexity.}
\label{lemma:amortization}
\end{lemma}

\textit{Proof.} As the metric tensor $g_{ij} \to 0$ along the path $\gamma$, the geodesic work term vanishes:$\lim_{t \to \infty} \int_{0}^{1} \sqrt{e^{-\lambda t} g_{ij} \dot{\gamma}^i \dot{\gamma}^j} \, dt = 0$. Simultaneously, as the distance shrinks, the exploratory volume accessible to the agent collapses, driving the entropy production $\mathcal{S}$ to zero (certainty). Therefore, the total action collapses to the residual cost of signal transmission:$\lim_{t \to \infty} \mathcal{A}(\gamma) = \mathcal{A}_{\text{retrieval}} = \epsilon$.

The above lemma proves that condensation creates a topological ``wormhole'' (realizing the conversion of
time into space): a complex reasoning chain that originally required exponential energy (search) is physically transformed into a zero-distance step (navigation), which we call ``topological condensation'' \cite{nakahara2018geometry}. Importantly, not every closed trajectory may be safely condensed: for condensation to yield a reusable scaffold rather than a spurious shortcut, the system must determine when a closed chain encodes a valid causal relation that can be properly amortized into structure.

\noindent\textbf{Generalized Hebbian Learning as Metric Contraction}
We propose that topological condensation is physically implemented via generalized Hebbian learning \cite{hebb1949organization}. While classical Hebbian theory posits that ``neurons that fire together, wire together'', the local rule lacks a mechanism to prevent runaway excitation. In our framework, we introduce a topological constraint: Hebbian plasticity is gated by cycle closure, which becomes \emph{Topological Hebb Rule}. Let $W_{ij}$ be the synaptic connectivity between two representation nodes. We interpret the weight not merely as gain, but as the inverse of the Riemannian metric distance $d_{ij}$ in the latent manifold \cite{muller1996hippocampus}: $d_{ij} \propto \frac{1}{W_{ij}}$.
The learning update is governed by the boundary operator: \begin{equation}
\Delta W_{ij} \propto 
\begin{cases}
\alpha \, (\text{Activity}_i \cdot \text{Activity}_j) 
    & \text{if } i,j \in \gamma \text{ and } \partial \gamma = 0, \\[6pt]
0   & \text{otherwise.}
\end{cases}
\end{equation}
The generalized Hebbian rule ensures that synaptic strengthening (metric contraction) occurs only when the neural activity forms a closed, self-consistent loop (e.g., sharp-wave ripple during sleep \cite{wilson1994reactivation}). Over repeated traversals of a closed cycle, the Hebbian integration drives $W \to W_{max}$, effectively driving the geodesic distance $d(\gamma) \to \epsilon$. Topologically, we are using a quotient map where distinct neurons participating in the cycle become functionally coincident. The path becomes a point, which we call ``\emph{manifold collapse}''. 

\subsection{The Parity Principle and Scaffold-Flow Dynamics}
\label{sec:parity_scaffold}
We propose that the brain relies on the fundamental topological invariant \cite{wheeler1990information}: ``The Boundary of a Boundary Vanishes'' \cite{hatcher2005algebraic}. In algebraic topology, a valid cycle (a loop with no ends) has a boundary of zero ($\partial \gamma = 0$). We map the closure principle directly to the cognitive process of prediction error minimization \cite{friston2010free}: Let $\vec{P}$ be the top-down prediction vector (Forward path) and $\vec{O}$ be the bottom-up sensory observation (Backward path); then the combination forms an open chain $c = \vec{P} \cup \vec{O}$.

\begin{proposition}[Homological Validation]

An inference is valid if and only if the chain $c$ forms a closed homological cycle, satisfying: $\partial c = \vec{P} - \vec{O} = 0$. 
\end{proposition}

If $\partial c \neq 0$, the cycle is broken. A non-zero boundary manifests physically as free energy or prediction error \cite{friston2010free}. The system must perform work (search) to deform the path until the boundary vanishes. Only when $\partial c = 0$ (cycle closure) can the structure be subjected to condensation, which establishes the rigorous constraint for the ``\textbf{Geometry of Certainty}'': certainty is the topological state where the boundary of the inference chain vanishes.
The geometric foundation of our framework is the conservation law $\partial^2 = 0$, stating that the boundary of a boundary is null \cite{wheeler1990information}, which governs the existence of topological features: a feature exists precisely where the boundary operator fails to close a chain (a cycle) that is not itself the boundary of a higher-dimensional object.

The global structure of a manifold $\mathcal{M}$ is summarized by two opposing metrics. First, the Euler characteristic $\chi$, which measures the net structural stability ($b_k=dim (H_k)$ is the $k$-th Betti number for the homology group $H_k$):
\begin{equation}
\chi(\mathcal{M}) = \sum_{k=0}^{n} (-1)^k b_k = \mathcal{H}_{\text{even}} - \mathcal{H}_{\text{odd}}
\end{equation}
Second, the homological capacity $C_H$, which measures the gross topological load (metabolic cost):
\begin{equation}
C_H(\mathcal{M}) = \sum_{k=0}^{n} b_k = \mathcal{H}_{\text{even}} + \mathcal{H}_{\text{odd}}
\end{equation}
We propose that the interplay between these metrics reflects a physical antagonism between two distinct regimes of information, formalized as the \textit{Homological Parity Principle}.

\begin{principle}[Homological Parity]
The topology of a cognitive system is partitioned into two conjugate phases based on dimension parity:
\begin{itemize}
    \item \textbf{Even Parity ($\mathcal{H}_{\text{even}}$)} represents the \textbf{Scaffold} ($\Phi$). These features (components $\beta_0$, cavities $\beta_2$) define static boundaries and invariant structures (condensed content).
    \item \textbf{Odd Parity ($\mathcal{H}_{\text{odd}}$)} represents the \textbf{Flow} ($\Psi$). These features (cycles $\beta_1$) define dynamic trajectories and recursive loops (active context).
\end{itemize}
\end{principle}

\noindent Parity-based partition implies a Variational Principle of Memory: The cortex evolves to maximize structural stability ($\chi$) while minimizing metabolic complexity ($C_H$).
\[
\begin{aligned}
\min (C_H - \chi)
&\;\;\implies\;\;
\min \!\left( (\mathcal{H}_{\text{even}} + \mathcal{H}_{\text{odd}})
              - (\mathcal{H}_{\text{even}} - \mathcal{H}_{\text{odd}}) \right) \\[4pt]
&\;\;\implies\;\; \min (2\,\mathcal{H}_{\text{odd}}).
\end{aligned}
\]
\noindent Such derivation reveals a fundamental law: computational efficiency is achieved solely by minimizing $\mathcal{H}_{\text{odd}}$. Therefore, \emph{topological condensation}, the collapse of dynamic cycles ($\beta_1$) into static nodes ($\beta_0$), is the optimal solution to the brain's thermodynamic constraints.

\begin{figure}[t]
\centering
\begin{tikzpicture}[
    >=Latex,
    font=\small,
    scaffold/.style={line width=1pt},
    flowloop/.style={line width=0.8pt, dashed},
    anchor/.style={circle, fill=black, inner sep=0pt, minimum size=2pt},
    lab/.style={font=\scriptsize, align=center}
]

\coordinate (S0) at (0,0);
\coordinate (S3) at (8,0);

\coordinate (A1) at ($(S0)!0.25!(S3)$);
\coordinate (A2) at ($(S0)!0.55!(S3)$);
\coordinate (A3) at ($(S0)!0.80!(S3)$);

\draw[scaffold] (S0) -- (S3);
\node[lab,below] at ($(S0)!0.5!(S3)$) {Scaffold $\sigma$ (stable content backbone)};


\draw[flowloop] 
  (A1) .. controls ($(A1)+(0,0.5)$) and ($(A1)+(-1.1,1.0)$) ..
  ($(A1)+(0,1.2)$) .. controls ($(A1)+(1.1,1.0)$) and ($(A1)+(0,0.5)$) .. (A1);
\node[lab,above] at ($(A1)+(0,1.4)$) {$\beta_{1}$};
\draw[->,flowloop] ($(A1)+(0,1.15)$) arc[start angle=90,end angle=-220,radius=1.0cm];

\draw[flowloop] 
  (A2) .. controls ($(A2)+(0,0.6)$) and ($(A2)+(-1.3,1.1)$) ..
  ($(A2)+(0,1.4)$) .. controls ($(A2)+(1.3,1.1)$) and ($(A2)+(0,0.6)$) .. (A2);
\node[lab,above] at ($(A2)+(0,1.6)$) {$\beta_{2}$};

\draw[flowloop] 
  (A3) .. controls ($(A3)+(0,0.5)$) and ($(A3)+(-1.0,1.0)$) ..
  ($(A3)+(0,1.2)$) .. controls ($(A3)+(1.0,1.0)$) and ($(A3)+(0,0.5)$) .. (A3);
\node[lab,above] at ($(A3)+(0,1.4)$) {$\beta_{3}$};

\draw[->,thick] ($(A1)+(0,-0.5)$) -- node[lab,below]{hypothesis $\gamma_i = \sigma + \sum_k a_{ik}\beta_k + \partial d_i$} ($(A3)+(0,-0.5)$);

\node[draw, rounded corners, inner sep=4pt, font=\scriptsize, align=left] 
  at ($(S0)!0.5!(S3)+(0,-2.0)$) {%
  \textbf{Scaffold--flow model.}\\
  $\sigma$: persistent content scaffold reused across hypotheses.\\
  $\beta_k$: contextual flow loops that deform $\sigma$ to explore\\
  \hspace*{1.4em}alternative interpretations or counterfactuals.\\
  Hypotheses $\gamma_i$ are realized as combinations of shared\\
  \hspace*{1.4em}scaffold plus task-dependent flows and small residuals $\partial d_i$.
};

\end{tikzpicture}
\caption{Scaffold-flow diagram for the homological memory model. 
The scaffold $\sigma$ forms a stable backbone of content, while 
loops $\beta_k$ represent contextual flows attached to it. 
Each hypothesis $\gamma_i$ is a deformation of the shared scaffold 
via a weighted combination of flows plus a small residual boundary.}
\label{fig:scaffold_flow}
\end{figure}
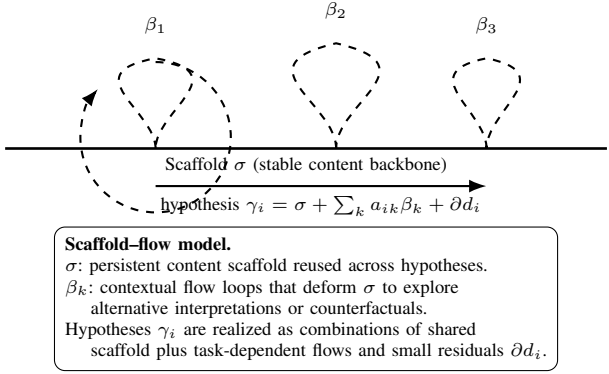

\noindent\textbf{The Scaffold-Flow Memory Model.} 
Projecting the parity principle onto biological architecture yields the scaffold-flow model. Memory is not a static store but a homological filtration process defined by the decomposition of a memory trace $\gamma_i$:
\begin{equation}
\gamma_i
= \underbrace{\sigma}_{\text{Scaffold } (\Phi)}
  \;+\;
    \underbrace{\sum_{k=1}^{b_i} a_{ik}\,\beta_k}_{\text{Flow } (\Psi)}
  \;+\;
    \underbrace{\partial d_i}_{\text{Noise}}
\label{eq:homological_memory_model}
\end{equation}

\begin{itemize}\item \textbf{Scaffold ($\Phi=\sigma$):} The stable backbone ($\partial \sigma=0$) common to multiple traces. It acts as a low-pass filter, preserving invariant structure (semantic memory \cite{tulving1972episodic}).\item \textbf{Flow ($\Psi$):} A linear combination of basis loops $\{\beta_k\}$ representing specific exploratory trajectories (episodic memory \cite{tulving1972episodic}).\item \textbf{Noise ($\partial d_i$):} The boundary term. By the law $\partial^2=0$, the noise component is topologically trivial and transient, representing sensory details discarded during consolidation.\end{itemize}

\noindent\textbf{Memory Consolidation as Topological Condensation.} The scaffold-flow model reinterprets memory consolidation not as data transfer, but as a topological phase transition implemented by generalized Hebbian learning. When a novel event (a high-fidelity $\Hodd$ cycle) is consolidated \cite{squire2015memory}, the system extracts its invariant structure. The specific temporal links are relaxed, and the persistent features are frozen into the $\Heven$ scaffold (e.g., $H_1$-cycle becomes a $H_0$-dot after condensation). With external sensory noise absent, $\Hodd$ content, dominated by replayed episodic memory traces \cite{wilson1994reactivation}, provides training data to anneal the $\Heven$ scaffold. 
Short-term traces, dominated by transient chains ($\partial d_i$) within the spatiotemporal complex $\mathcal{K}_\delta(t)$, correspond to high-entropy sensory noise. Through recurrent replay, these transient boundaries are annihilated ($\partial d_i \to 0$) via the cycle-closure principle, leaving behind only the persistent chains ($\sigma + \sum a_{ik}\beta_k$) that form the stable memory manifold. 

Through recurrent replay and synchronization \cite{lisman2013theta}, these short-term traces undergo topological averaging, progressively stripping away non-recurrent boundaries ($\partial d_i \to 0$) to reveal the invariant loops that satisfy $ \partial^2 = 0 $. The consolidation process refines the $\mathcal{H}_{\text{even}}$ scaffold ($\sigma$) through homological promotion. If a specific contextual loop $\beta_k$ persists across multiple disjoint episodes, the system condenses it into the backbone itself ($\sigma_{new} \leftarrow \sigma_{old} \cup \text{Condense}(\beta_k)$). Under the structure-before-specificity (SbS) framework, consolidation amounts to a dual operation: 1) filtration - converting transient boundaries ($ \gamma \in \operatorname{im}\partial_2 $) into persistent homology classes ($ [\gamma] \in H_1(\mathcal{K}_\delta) $) \cite{edelsbrunner2008persistent}; 2) deformation - embedding high-entropy contextual flows ($ \Psi=\Hodd$) that survive filtration into the low-entropy structural scaffold ($ \Phi=\Heven $). The resulting long-term memory (LTM) trace is not a static replica of the original event but a structural attractor, retrievable through the completion of a homological loop rather than the reactivation of an exact firing pattern.

\begin{figure}[h]
\centering
\resizebox{\columnwidth}{!}{
\begin{tikzpicture}[
    module/.style={draw, thick, rounded corners, minimum width=3.6cm, minimum height=1.4cm, align=center},
    arrow/.style={->, thick},
    dashedarrow/.style={->, thick, dashed},
    font=\small
]

\node[module, fill=blue!10] (context) at (0, 0) {Context \\ \( \Psi_t \)};
\node[module, fill=green!10] (retrieve) at (4.5, -2.5) {Retrieval \\ \( \hat{\Phi}_t = \mathcal{R}(\Phi_{t+1}, \Psi_t) \)};
\node[module, fill=orange!10] (adapt) at (4.5, 0) {Bootstrapping \\ \( \Phi_t = \mathcal{F}(\hat{\Phi}_t, \Psi_t) \)};
\node[module] (predict) at (9, 0) {Predictive Update \\ \( \Phi_{t+1} \)};

\draw[arrow] (context.east) -- ++(0.5, 0) |- (retrieve.west);
\draw[arrow] (retrieve.north) -- (adapt.south);
\draw[arrow] (adapt.east) -- (predict.west);
\draw[dashedarrow] (predict.south) -- ++(0, -1.5) node[midway, right] {\small reuse} |- (retrieve.east);

\node at (4.5, 1.3) {\textbf{Memory-Amortized Inference Cycle}};
\node at (7.8, -2.9) {\(\mathcal{M} = \{ (\Psi^{(i)}, \Phi^{(i)}) \}\)};

\end{tikzpicture}
}
\caption{Cycle of MAI. Instead of recomputing \(\Phi^* = \arg\min \mathcal{L}(\Psi, \Phi)\), the system reuses prior trajectories: \(\Phi_{t+1}\) and \(\Psi_t\) guide memory-based retrieval via \(\mathcal{R}\), and bootstrapping \(\mathcal{F}\) updates the latent state \(\Phi_t\). The process forms a self-consistent loop grounded in structured memory.}
\vspace{-0.2in}
\label{fig:mai-cycle}
\end{figure}
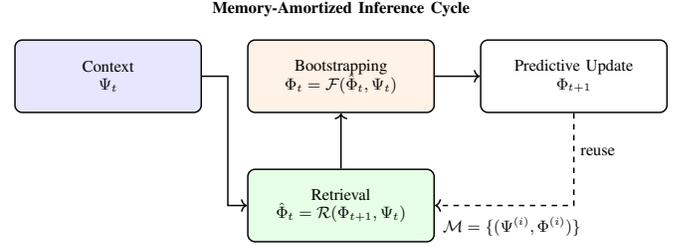

\section{Memory-Amortized Inference (MAI): Topological Trinity and Dual-Mode Operations}
\label{sec:MAI_trinity}

How does a cognitive system physically minimize the thermodynamic action defined in Section \ref{sec:theoretical_foundations}? We propose that the brain implements a specific class of computation we term MAI. MAI solves the search problem by replacing the expensive \emph{de novo} optimization of a likelihood function with a cheaper cycle of retrieval and adaptation (Fig. \ref{fig:mai-cycle}). During amortized \textit{inference}, a topological expansion cycle where the $\Heven$ scaffold provides top-down predictions to constrain the $\Hodd$ flow. The inference mode directly formalizes hierarchical Bayesian inference \cite{rao1999predictive}, with perception as the convergence of the cycle, where the $\Hodd$ flow ``snaps'' to the nearest low-energy state on the $\Heven$ scaffold. 

\subsection{Memory-Amortized Inference via the Topological Trinity}

We start with a formal definition of MAI next.

\begin{definition}[Memory-Amortized Inference]Let $ \mathcal{M} = \{ (\Psi^{(i)}, \Phi^{(i)}) \}_{i=1}^N $ be a memory of prior context–content pairs, and let $ \mathcal{R}: \mathcal{X} \times \mathcal{M} \to \mathcal{S} $ be a retrieval-and-adaptation operator and $\mathcal{F}: \mathcal{S}\times\mathcal{X}\to\mathcal{S}$ be the bootstrapping update operator implemented via generative simulation. Inference is said to be \emph{memory-amortized} if it is formulated as a structural cycle between \emph{content} $ \Phi $ and \emph{context} $ \Psi $, where memory acts as a reusable substrate for inference:
\begin{equation}\Phi_{t+1} = \mathcal{F}(\Phi_t, \Psi_t), \quad \Phi_t \approx \mathcal{R}(\Phi_{t+1}, \Psi_t) \label{eq:MAI} \end{equation}
in lieu of directly optimizing $ \Phi^* $, such that the expected cost satisfies $\mathbb{E}{\Psi} \left[ \mathcal{L}(\Psi, \hat{\Phi}) \right] \leq \mathbb{E}{\Psi} \left[ \mathcal{L}(\Psi, \Phi^*) \right] + \varepsilon$, for some amortization gap $ \varepsilon \ll \mathcal{L}(\Psi, \cdot) $, and where the runtime cost of $ \mathcal{R} $ is substantially lower than full inference.\end{definition}

The above definition implies that inference is not a feedforward path but a closed loop involving two conjugate forces: the forward generative bootstrapping ($\mathcal{F}$) and the backward retrieval constraint ($\mathcal{R}$). We propose that the \textbf{Topological Trinity}: \[ \emph{Search $\to$ Closure $\to$ Condensation}\] is the physical algorithm that executes these operators to minimize the amortization gap $\varepsilon$.

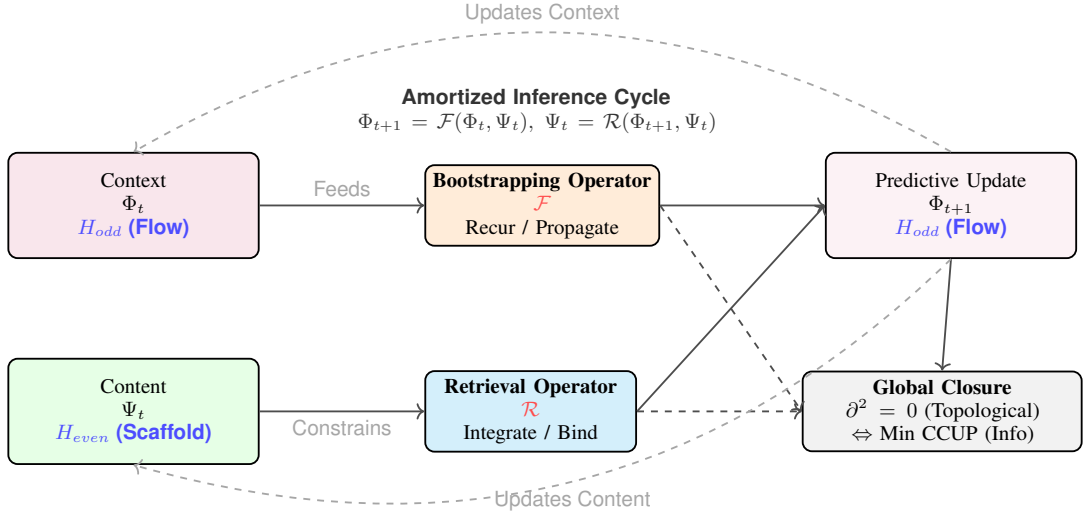
\begin{figure*}[h]
    \centering
    \resizebox{0.8\linewidth}{!}{
\begin{tikzpicture}[
    node distance=1.5cm and 2.5cm, 
    module/.style={draw, thick, rounded corners, minimum width=3.8cm, minimum height=1.6cm, align=center},
    op/.style={draw, thick, rounded corners, minimum width=3.2cm, minimum height=1.0cm, align=center, fill=gray!10},
    homology_box/.style={draw, thick, rounded corners, minimum width=3.0cm, minimum height=1.2cm, align=center, fill=blue!5},
    arrow/.style={->, thick, draw=black!70},
    dashedarrow/.style={->, thick, dashed, draw=gray!70},
    label_style/.style={font=\sffamily\small, text=gray!70},
    cycle_label/.style={font=\sffamily\bfseries\small, text=black!80},
    font=\small
]

\node[module, fill=purple!10] (phi_content) at (0, 0) {
    Context \\ \(\Phi_t\) \\
    \textcolor{blue!70}{\sffamily\bfseries\small $H_{odd}$ (Flow)}
};

\node[module, fill=green!10, below=of phi_content] (psi_context) {
    Content \\ \(\Psi_t\) \\
    \textcolor{blue!70}{\sffamily\bfseries\small $H_{even}$ (Scaffold)}
};

\node[op, fill=orange!15, right=of phi_content] (bootstrap_op) {
    \textbf{Bootstrapping Operator} \\
    \textcolor{red!70}{\sffamily\bfseries\small \(\mathcal{F}\)} \\
    Recur / Propagate
};

\node[op, fill=cyan!15, right=of psi_context] (retrieval_op) {
    \textbf{Retrieval Operator} \\ 
    \textcolor{red!70}{\sffamily\bfseries\small \(\mathcal{R}\)} \\
    Integrate / Bind
};

\node[module, fill=purple!5, right=of bootstrap_op] (phi_next) {
    Predictive Update \\ \(\Phi_{t+1}\) \\
    \textcolor{blue!70}{\sffamily\bfseries\small $H_{odd}$ (Flow)}
};

\node[homology_box, fill=gray!10, text width=4cm, right=of retrieval_op] (closure_goal) {
    \textbf{Global Closure} \\ \(\partial^2=0\) (Topological) \\
    $\Leftrightarrow$ Min CCUP (Info)
};

\draw[arrow] (phi_content.east) -- node[above, label_style] {Feeds} (bootstrap_op.west);
\draw[arrow] (psi_context.east) -- node[below, label_style] {Constrains} (retrieval_op.west);

\draw[arrow] (bootstrap_op.east) -- (phi_next.west);
\draw[arrow] (retrieval_op.east) -- (phi_next.west);

\draw[arrow] (phi_next.south) -- (closure_goal.north);

\draw[arrow, dashed] (bootstrap_op.east) -- (closure_goal.west);
\draw[arrow, dashed] (retrieval_op.east) -- (closure_goal.west);

\draw[dashedarrow, bend right=30] (phi_next.north) to node[above, label_style] {Updates Context} (phi_content.north);
\draw[dashedarrow, bend left=30] (phi_next.south) to node[below, label_style] {Updates Content} (psi_context.south);

\node[cycle_label, text width=6cm, align=center] at ($(bootstrap_op)!0.5!(retrieval_op) + (0, 3.cm)$) {
    \textbf{Amortized Inference Cycle} \\
    $\Phi_{t+1}=\mathcal{F}(\Phi_t,\Psi_t),~ \Psi_t=\mathcal{R}(\Phi_{t+1},\Psi_t)$
};

\end{tikzpicture}
}
    \caption{\textbf{Memory-Amortized Inference (MAI) Cycle as a Homological Parity Split}.
The computational model for minimizing content-context uncertainty. The cycle is partitioned by topological parity. Odd-dimensional homology ($H_{odd}$), representing content ($\Phi_t$), is processed by the bootstrapping operator ($\mathcal{F}$). Simultaneously, even-dimensional homology ($H_{even}$), representing context ($\Psi_t$), is processed by the retrieval operator ($\mathcal{R}$). Both operators drive the predictive update ($\Phi_{t+1}$), which in turn updates the system, seeking a state of global closure ($\partial^2=0$).}
\vspace{-0.5cm}
    \label{fig:MAI}
\end{figure*}

\noindent\textbf{Phase I: Search (The Retrieval Operator $\mathcal{R}$)} The cycle begins with search, which corresponds physically to the execution of the retrieval operator $\mathcal{R}$. When a novel context $\Psi_{new}$ is encountered, the system cannot immediately compute the optimal content $\Phi^*$. Instead, it queries the memory manifold $\mathcal{M}$ to retrieve a prior content $\Phi_t$ that is topologically homologous to the current input: $\Phi_{prior} = \mathcal{R}(\Psi_{new}, \mathcal{M})$. The first phase represents the system's backward look into its history. It is the \emph{adaptation} step where the system identifies a candidate scaffold. In the language of the scaffold-flow model, phase-I is the activation of a candidate $\Hodd$ flow from the latent space. In computational complexity isomorphism, phase-I is Savitch's recursion (trading time to save space), where nondeterministic problems in deterministic space are solved by recursively searching for midpoints \cite{savitch1970relationships}.

\noindent\textbf{Phase II: Closure (The Bootstrapping Cycle $\mathcal{F} \leftrightarrow \mathcal{R}$)} Retrieval alone is insufficient; the retrieved memory must be validated against the current reality (i.e., satisfying Bellman Equation \cite{bellman1966dynamic}). The validation is the closure phase, generated by the interaction between the bootstrapping operator $\mathcal{F}$ and the retrieval $\mathcal{R}$. The system uses the retrieved prior to \emph{bootstrap} a simulation of the present: $\hat{\Phi}_{t+1} = \mathcal{F}(\Phi_{prior}, \Psi_{new})$. Note that MAI requires that the simulation be consistent with the retrieval, forming the structural cycle defined in Eq. \eqref{eq:MAI}: $\Phi_t \approx \mathcal{R}(\mathcal{F}(\Phi_t))$. Such a closure requirement implies that the forward prediction (bootstrapping) and the backward constraint (retrieval) must meet to form a closed homological loop ($\partial \gamma = 0$). If $\mathcal{F}$ diverges from $\mathcal{R}$, the cycle is open ($\partial \gamma \neq 0$), generating prediction error \cite{friston2010free}. If $\mathcal{F}$ aligns with $\mathcal{R}$, the cycle closes, achieving homological stability via generalized Hebbian learning, as we define next.

\begin{definition}[Homological Stability]
A cognitive state is \emph{stable} if and only if the neural activation pattern forms a non-trivial homology class $[\gamma] \in H_1(\mathcal{M})$. Such topological closure acts as the system's internal truth test. An open chain ($\partial \gamma \neq 0$) represents uncertainty or error (free energy); a closed cycle represents a validated semantic unit.
\end{definition}
    
\noindent\textbf{Phase III: Condensation (Minimizing the Cost)}
The final phase, topological condensation implementing memory consolidation \cite{squire2015memory}, is the mechanism of memoization that ensures the runtime cost is substantially lower, as required by the objective of amortization.
Once the cycle $\mathcal{F} \leftrightarrow \mathcal{R}$ is stable, the system performs a quotient map operation ($q$), collapsing the entire computational loop into a single addressable vertex:
$\text{Cost}(\mathcal{R}) \to \epsilon$ (generalization of Hebbian learning postulate to its asymptoic limit \cite{hebb1949organization}).
By condensing the verified cycle, the system updates the memory $\mathcal{M}$. The complex structural cycle between $\Phi$ and $\Psi$ is amortized into a static link. Future invocations of $\mathcal{R}$ for the same context no longer require the expensive simulation loop $\mathcal{F}$; they simply access the condensed vertex. This is the physical realization of the inequality $\varepsilon \ll \mathcal{L}(\Psi, \cdot)$ by executing the memoization step (i.e., caching the topological result). Together, the trinity transformation allows the cortex to solve NPSPACE problems using polynomial resources based on Savitch's theorem (search), Bellman equation (closure), and dynamic programming (condensation).

\noindent\textbf{A Concrete Example: The Commute}
To ground these topological abstractions, consider the everyday cognitive task of learning a route to a new workplace.

\begin{enumerate}
\item \textbf{Search (Scaffold-Building):}
Initially, the city is a set of disconnected landmarks (high $\beta_0$) and unknown constraints. The driver must explore, testing various streets. The manifold is fragmented with high-entropy context.

\item \textbf{Closure (Odd Flow $\beta_1$):} 
Once a successful route is found, it exists as a \textbf{sequence of actions}: ``Turn left at the elm tree, drive two miles, turn right at the gas station.'' Topologically, it is a \textbf{1-cycle} ($\mathcal{H}_{odd}$), a trajectory that must be actively traversed in time. It is a \textit{dynamic loop} that consumes metabolic energy and attention.

\item \textbf{Condensation (Even Structure $\beta_0$):} 
Over time, the brain amortizes the learned path. The complex sequence collapses. The entire trajectory ``Home $\to$ Work'' is compressed into a single conceptual unit: The Commute. 
Topologically, the \textbf{Odd cycle ($\beta_1$)} at level $k$ has condensed into an \textbf{Even node ($\beta_0$)} at level $k+1$.

\end{enumerate}

\noindent\textbf{The Result:} The driver no longer simulates the turns but simply navigates the meta-scaffold. They can now reason at a higher level of abstraction: ``I will do \textit{The Commute}, then \textit{Go to Gym}.'' The brain has traded the \textbf{time} required to simulate the path for the \textbf{space} required to store the node, satisfying the variational principle of minimizing active flow ($\beta_1$).

\subsection{Parity Alternation: The Trinity as a Mode Switch.}
The topological trinity serves as the temporal clock that implements the parity alternation required by the scaffold-flow model (Fig. \ref{fig:MAI}). It temporally segregates the optimization of context and content to avoid catastrophic interference \cite{wang2024comprehensive}.

\noindent\textbf{1. Inference Mode (Wake: Search $\to$ Closure).} The waking mode is a context-before-content cycle that formalizes hierarchical Bayesian inference \cite{rao1999predictive}. The wake mode consists of the search and closure phases in trinity transformation: the system fixes the scaffold $\Phi$ (using it as the constraint $\mathcal{R}$) and optimizes the flow $\Psi$ (via bootstrapping $\mathcal{F}$). The $\Heven$ scaffold provides top-down predictions to constrain the $\Hodd$ flow, with perception occurring as the convergence of the cycle where the flow snaps to the nearest low-energy state on the scaffold. The goal is to find a $\Psi$ that best fits the existing $\Phi$.

\noindent\textbf{2. Learning Mode (Sleep: Condensation).} The sleep mode is an inverted Structure-before-Specificity (SbS) process \cite{buzsaki1996hippocampo}. In this phase, the system fixes the valid flow $\Psi$ (the closed cycle) and optimizes the scaffold $\Phi$ (via metric contraction) as we have discussed in Sec. \ref{sec:parity_scaffold}. With external sensory noise absent, the $\Hodd$ content, dominated by replayed episodic memory traces \cite{wilson1994reactivation}, provides the training data to anneal the $\Heven$ scaffold. The goal is to deform $\Phi$ to permanently accommodate $\Psi$.

Together, the dual-mode operation based on the trinity transformation creates a rhythmic ticktock dynamics of cognition: \emph{inference} (finding the path) is immediately followed by \emph{learning} (paving the path), ensuring that the system never updates its structure based on unverified flows.

\noindent\textbf{Isomorphism to EM and Wake-Sleep Algorithms}
The parity alternation dynamics described above reveal that the topological trinity is structurally isomorphic to the celebrated Expectation-Maximization (EM) algorithm \cite{dempster1977maximum}, implemented physically via a Wake-Sleep architecture \cite{hinton1995wake}. We posit that MAI can be interpreted as a Topological EM that alternates between:
1) \textbf{E-Step (Inference Mode):} The system fixes the scaffold ($\Phi$) and optimizes the flow ($\Psi$), which corresponds to the search and closure phases, where the agent minimizes the homological boundary ($\partial \gamma \to 0$) to find the valid geodesic for the current input; 2) \textbf{M-Step (Learning Mode):} The system fixes the flow ($\Psi$) and updates the scaffold ($\Phi$), which corresponds to the condensation phase. During the condensation, the agent minimizes the metric distance ($d \to 0$) along the validated path, effectively maximizing the topological likelihood of the memory.
We also note the connection between MAI and Hinton's wake-sleep algorithm \cite{hinton1995wake}. Because simultaneous optimization of flow and scaffold is computationally intractable (e.g., the moving target problem \cite{mnih2015human}), biological systems temporally segregate these operations. Wake mode corresponds to the topological E-Step: the manifold is held rigid to support high-fidelity inference flows ($\beta_1$). Sleep mode matches the topological M-Step: the sensory input is disconnected, and the system replays successful cycles to permit aggressive metric contraction ($\beta_1 \to \beta_0$). Our interpretation aligns with the Synaptic Homeostasis Hypothesis \cite{tononi2014sleep}: sleep is the thermodynamic phase transition where the high-energy costs of wakeful Search are amortized into the low-energy structure of memory.

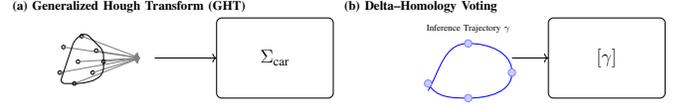
\begin{figure}[h]
\centering
\resizebox{\linewidth}{!}{
\begin{tikzpicture}[
    edgept/.style={circle, draw=black, fill=gray!20, inner sep=1pt},
    latentpt/.style={circle, draw=blue!50, fill=blue!20, inner sep=2pt},
    accumulator/.style={rounded corners, draw=black, thick, minimum width=3.2cm, minimum height=2.2cm},
    arrow/.style={->, thick},
    caroutline/.style={draw=black, thick, rounded corners},
    vote/.style={->, thick, draw=gray},
    loop/.style={thick, draw=blue},
    label/.style={font=\small}
]

\node at (0,3.4) {\textbf{(a) Generalized Hough Transform (GHT)}};

\foreach \x/\y in {-1.8/2.3, -1.3/2.6, -0.9/2.2, -1.0/1.6, -1.5/1.3, -1.9/1.6, -1.4/1.9, -0.7/1.9} {
    \node[edgept] at (\x,\y) {};
    \draw[vote] (\x,\y) -- (0.3,2); 
}

\draw[caroutline] (-1.9,1.3) -- (-1.5,2.5) -- (-1.2,2.7) -- (-0.9,2.4) -- (-0.7,2.2) -- (-0.7,1.6)
 -- (-0.9,1.4) -- (-1.3,1.3) -- cycle;

\draw[arrow] (0.7,2) -- (2.4,2);

\node[accumulator, label=below:{\small Template Center Accumulator}] at (4,2) {};
\node at (4,2) {\Large $\Sigma_{\text{car}}$};

\node at (8.,3.4) {\textbf{(b) Delta–Homology Voting}};

\draw[loop] (8.2,1.1) to[out=60,in=180] (9.3,2.4)
            to[out=0,in=90] (10.5,1.6)
            to[out=270,in=0] (9.3,0.9)
            to[out=180,in=300] (8.2,1.3);

\node at (9.3,2.8) {\scriptsize Inference Trajectory \( \gamma \)};

\foreach \x/\y in {8.2/1.3, 9.3/2.4, 10.5/1.6, 9.3/0.9} {
    \node[latentpt] at (\x,\y) {};
}

\draw[arrow] (10.5,2) -- (11.5,2);

\node[accumulator, label=below:{\small Homology Classifier}] at (13.1,2) {};
\node at (13.1,2) {\Large $[\gamma]$};

\end{tikzpicture}
}
\caption{Comparison of (a) Generalized Hough Transform (GHT) for generic shape detection and (b) Delta-Homology Voting in latent cycle-closure inference. In GHT, edge features vote toward a known geometric template (e.g., a car) via a reference point accumulator. In delta–homology, each inference step emits a delta-like activation along a latent cycle \( \gamma \), which triggers a persistent memory only when the full topological loop is reconstructed.}
\label{fig:delta_ght_car}
\end{figure}

\noindent\textbf{Consistency with the Delta-Homology Analogy}
The homological parity principle provides the rigorous topological justification for the \emph{Delta-Homology Analogy}, a topological extension of generalized Hough transform (GHT) \cite{ballard1981generalizing} (Fig. \ref{fig:delta_ght_car}). A frequent conceptual challenge is reconciling the dynamic nature of memory flow (e.g., a loop, $H_1$) with its static representation as a memory trace (a delta function, $\delta$). We resolve the problem via the topology of the \emph{moduli space}.
\begin{enumerate}\item \textit{The Dynamic View (Odd):} During the \emph{bootstrapping} phase, the system generates a trajectory. To stabilize, the trajectory must form a closed loop, establishing a generator $[\gamma] \in H_1(\mathcal{K})$, which is an odd-dimensional entity representing flow.\item \textit{The Storage View (Even):} During the \emph{condensation} phase, the system amortizes the formed loop. We treat the entire homology class $[\gamma]$ as a single point in the moduli space of cycles. The probability mass collapses onto this invariant, forming a Dirac delta distribution $\delta_\gamma$ (i.e., manifold collapse).
\end{enumerate}
In summary, the Delta is simply the \textit{even-dimensional projection} of an \textit{odd-dimensional cycle}, confirming the trinity transformation: \[\underbrace{\text{Search}}_{\text{Even Scaffold}} \xrightarrow{\text{generates}} \underbrace{\text{Closure}}_{\text{Odd Cycle } (H_1)} \xrightarrow{\text{condenses to}} \underbrace{\text{Synthesis}}_{\text{Even Trace } (\delta \in H_0)}\] The ``Delta'' is the static handle by which the brain grabs the dynamic loop.

\section{Recursive Condensation and Biological Implementation}
\label{sec:recursive_condensation}

The topological trinity describes the lifecycle of a single cognitive feature. However, the world is not flat; it is causally deep. Atoms form molecules, molecules form materials, materials form objects, and objects form scenes (``More is Different'' \cite{anderson1972more}). The parity principle dictates that the brain must mirror the depth of nested structures in nature.
We propose that the cortex achieves the mirroring by applying the condensation operator $\Phi$ recursively, creating a ``Tower of Scaffolds'' where the ceiling of one level becomes the floor of the next \cite{simon1962architecture}.

\subsection{Recursive Condensation: Building the Tower of Scaffolds}

\noindent\textbf{From Flow to Scaffold}
In standard neural network theory, layers are often viewed as performing identical filtering operations at different scales. In our framework, the relationship between layers is transformational.
The output of layer $L_k$ is not merely a feature map but a topological phase transition.
\begin{itemize}
    \item At Level $k$ (Flow / $\Psi$): A pattern exists as a temporal correlation - i.e., a rapid sequence of firing (a cycle $\beta_1$). The system must expend energy to maintain this relationship.
    \item At Level $k+1$ (Scaffold / $\Phi$): That same pattern exists as a spatial address such as a single neuron or microcolumn (a vertex $\beta_0$). The system treats it as a static atom.
\end{itemize}

We formalize a level-to-level transition as the conversion of odd-dimensional homology (cycles) into even-dimensional homology (components).

\begin{proposition}[Recursive Condensation]
\label{prop:recursion}
The topological structure of the cortical hierarchy is generated by the recursive transformation of homological flow into structural scaffold:
$\mathcal{H}_{\text{odd}}^{(k)} \xrightarrow{\text{Condense}} \mathcal{H}_{\text{even}}^{(k+1)}$
Specifically, a stable, high-frequency limit cycle ($\beta_1$) at level $k$ is treated as a static atomic unit ($\beta_0$) at level $k+1$.
\end{proposition}

\noindent\textbf{Proof (Sketch).}
Let $C_k$ be a stable cycle in the manifold of layer $k$. The Condensation Operator $q$ identifies all points $x \in C_k$ as equivalent ($x \sim x'$).
The quotient space $\mathcal{M}_k / \sim$ effectively contracts $C_k$ to a point $p$.
The point $p$ becomes a basis element for the manifold of layer $k+1$, $\mathcal{M}_{k+1}$.
What was a path with non-zero length and non-zero time cost in $L_k$ becomes a point with zero length and zero time cost in $L_{k+1}$.

\noindent\textbf{The Tower of Scaffolds}
The above recursion solves the combinatorial explosion of high-level reasoning.
Without recursion, representing a concept like Democracy would require activating the chain of all human interactions that define it, an impossible computational load (Minsky's Search Problem \cite{minsky1967computation}).
With Recursive Condensation, the brain builds a Tower of Scaffolds for visual perception \cite{dicarlo2012does}, as an example:
\begin{enumerate}
    \item V1 (Pixels $\to$ Edges): Correlations of light ($\beta_1$) condense into line segments ($\beta_0$).
    \item V4 (Edges $\to$ Shapes): Closed cycles of lines ($\beta_1$) condense into geometric primitives ($\beta_0$).
    \item IT (Shapes $\to$ Objects): Configurations of shapes ($\beta_1$) condense into semantic entities (e.g., ``Cat'') ($\beta_0$).
    \item PFC (Objects $\to$ Concepts): Causal loops between entities ($\beta_1$) condense into abstract rules ($\beta_0$).
\end{enumerate}

By the time the signal reaches the prefrontal cortex (PFC) \cite{fuster2008prefrontal}, the atoms of computation are no longer sensory details but entire causal histories condensed into single points based on the delta-homology analogy. Recursive condensation allows the agent to manipulate complex narratives with the same low-latency efficiency as it manipulates edges, effectively achieving zero-shot generalization \cite{brown2020language} by navigating the highest level of the scaffold.
The framework of recursive condensation offers a topological explanation for why deep learning works \cite{ayzenberg2025sheaf}. Depth is not about parameter count; it is about topological resolution.
A shallow network such as a multi-layer perception (MLP) \cite{rosenblatt1958perceptron} tries to map input $\to$ output in one condensation step. If the causal depth of the problem exceeds the depth of the network, the ``wormhole'' required is too long and unstable (high error).
A deep network builds intermediate stable bridges (i.e., scaffolds), which solves the problem by breaking the long geodesic into a series of short, verifiable hops ($\beta_1 \to \beta_0 \to \beta_1 \to \beta_0$). Intelligence is the height of the scaffold one can maintain.

\subsection{Biological Implementation via Cortical Columns}\label{sec:biology}

The Topological Trinity and MAI are not merely computational abstractions; they describe the physical dynamics of the mammalian neocortex. We propose that the cortical column, the fundamental modular unit of the cortex, is the physiological operator that executes the Recursive Condensation defined in Proposition \ref{prop:recursion}. Our framework vindicates Mountcastle's Universal Algorithm Hypothesis \cite{mountcastle1957modality}: the uniformity of cortical tissue across modalities (vision, audition, motor) exists because every region performs the same topological operation: transforming a high-entropy input flow into a low-entropy condensed scaffold.

\noindent\textbf{Laminar Instantiation of the Trinity}
We map the three phases of the trinity transformation to the canonical laminar microcircuit of the cortical column \cite{douglas1989canonical}. The column acts as a vertical integration unit that receives search queries and emits condensed symbols.
\begin{enumerate}
    \item{Layer IV: The Search Interface ($\Psi_{in}$).}The input layer (Layer IV) receives high-dimensional sensory feedforward data (from Thalamus or lower cortical areas). In our framework, this constitutes the search phase initiation. The incoming spikes represent the context $\Psi_t$—a set of disconnected features ($\beta_0$ from below) that induce a high-frequency flow ($\beta_1$) within the local column.

    \item{Layers II/III $\leftrightarrow$ V/VI: The Closure Loop ($\mathcal{R} \leftrightarrow \mathcal{F}$).}The core computation occurs in the recurrent loop between the supragranular layers (II/III) and infragranular layers (V/VI). Bootstrapping ($\mathcal{F}$) is implemented by deep pyramidal neurons (Layer V) that project predictions upwards and outwards, representing the generative hypothesis (Content $\Phi$); Retrieval ($\mathcal{R}$) is implemented by superficial neurons (Layer II/III) that integrate these predictions with lateral context and feedforward error. Such a vertical oscillation constitutes the closure phase of the trinity transformation. The column attempts to achieve resonance where the descending prediction matches the ascending input ($\partial \gamma = 0$) \cite{larkum2013cellular}. Physiologically, the bidirectional resonance is observed as the synchronization of local field potentials (gamma oscillations) \cite{buzsaki2006rhythms}. An open cycle manifests as desynchronization and error signaling; a closed cycle manifests as stable resonance. 
    
    \item{Deep Pyramidal Bursting: Condensation ($\Phi_{out}$).} Once the internal loop stabilizes via cycle closure, the column emits a high-frequency burst via the large pyramidal neurons in Layer V/VI to downstream areas or higher cortical levels with output being the condensation. The complex internal processing of the column is collapsed into a single scalar frequency or spike packet. To the next layer in the hierarchy, this entire column becomes a single vertex in the manifold (i.e., manifold collapse for recursive condensation): \begin{equation}\text{Column}_k(\Psi_{flow}) \xrightarrow{\text{Condense}} \text{Neuron}_{k+1}(\Phi_{scaffold})\end{equation} 
\end{enumerate}

\noindent\textbf{Hebbian Learning as Metric Contraction} Under the framework of recursive condensation, we re-derive synaptic plasticity within the geometric framework. Let $W_{ij}$ be the synaptic connectivity between two representation nodes. As mentioned in Sec. \ref{sec:parity_scaffold}, the synaptic weight is inversely proportional to the Riemannian metric distance $d_{ij}$ in the latent manifold \cite{muller1996hippocampus}: $d_{ij} \propto \frac{1}{W_{ij}}$. When a column achieves Closure ($\partial \gamma = 0$), the neurons involved are active simultaneously. The resulting Long-Term Potentiation (LTP) increases $W_{ij}$, which topologically is equivalent to metric contraction. The system physically shortens the distance between the concepts involved in the successful cycle. In summary, biological learning is the process of warping the manifold to bring causally related states closer together, and expertise is the state where the manifold is so heavily folded that the distance between a problem and its solution approaches zero ($d \to \epsilon$), allowing for instant intuition (amortization) rather than exhaustive search.
Finally, we note that the above implementation unifies MAI with Friston’s Free Energy Principle \cite{friston2010free}. In our view, Prediction Error is simply the magnitude of the homological boundary ($\partial c \neq 0$). The cortical column minimizes Free Energy by performing work (Search) to deform the path until the boundary vanishes (Closure), at which point the energy is amortized into structure (Condensation).

\section{The Limits of Inference: Generalization and Hallucination}
\label{sec:limits}

The \emph{Amortization Efficiency Lemma} (Lemma \ref{lemma:amortization}) guarantees that topological condensation reduces the computational cost of inference to zero. However, the efficiency is not free. By warping the latent manifold to create wormholes between concepts, the system fundamentally alters the topology of the solution space.
We propose that generalization \cite{geman1992neural} and hallucination \cite{ji2023survey} are the positive and negative manifestations of the same geometric phenomenon: \emph{manifold folding} \cite{li2023toward}. The brain does not distinguish between them procedurally; both are the inevitable result of maximizing navigational speed.

\noindent\textbf{Manifold Folding and the Geometry of Shortcuts}
As discussed above, Hebbian learning drives metric contraction ($d_{ij} \to 0$). Over time, the contraction forces the high-dimensional manifold to fold or crumple, bringing distinct regions into proximity. Consider two concepts, $A$ and $B$, which are distant in the original, flat metric space ($d(A,B) \gg 0$). The energy-consuming search (to move from $A$ to $B$) requires traversing a long geodesic path (deduction); by contrast, folding is much more energy efficient \cite{buzsaki2006rhythms}: if $A$ and $B$ are frequently co-active, the metric between them collapses. The manifold folds such that $A$ and $B$ become neighbors, which creates a topological shortcut or wormhole (for faster navigation). The agent can now transition from $A$ to $B$ without traversing the intermediate logical steps (i.e., the mechanism of \emph{Intuition} or \emph{Insight} \cite{bowden2005new}).

\noindent\textbf{The Divergence of Utility}
The folded manifold $\mathcal{M}'$ allows for rapid navigation, but does it preserve truth? We distinguish two outcomes based on the homological parity between the folded manifold and the external world $\mathcal{W}$: 

\emph{1. Generalization (Homotopy Equivalence).}
Generalization occurs when the metric contraction respects the topological invariants of the environment.
Let $C_{sparrow}$ and $C_{robin}$ be two distinct cycles. If they share a common sub-chain (e.g., ``Wings''), the system contracts the distance between them.
If $d(C_{sparrow}, C_{robin}) < \epsilon$, the system treats them as the same class.
Such a contraction is valid if the world also treats them as equivalent for the task at hand.
\begin{equation}
\text{Generalization}: \quad \mathcal{M}' \simeq \mathcal{W} \quad (\text{Homotopy Equivalent})
\end{equation}
The system successfully applies a known solution (Sparrow logic) to a new problem (Robin context) because the fold in the manifold correctly reflects the causal structure of reality.

\emph{2. Hallucination (Singular Collapse).}
Hallucination occurs when the metric contraction violates the topological distinctions of the environment.
Consider Elon Musk ($C_1$) and Steve Jobs ($C_2$). Both share the invariant Tech CEO.
If the system over-condenses the invariant, the metric distance $d(C_1, C_2)$ may collapse to zero. The manifold develops a singularity where distinct entities merge.
\begin{equation}
\text{Hallucination}: \quad \mathcal{M}' \not\simeq \mathcal{W} \quad (\text{Topological Defect})
\end{equation}
When the system attempts to retrieve ``Founder of Apple,'' it navigates to the collapsed vertex. Because $C_1$ and $C_2$ are indistinguishable in $\mathcal{M}'$, the system may probabilistically traverse the path to Musk with high confidence.
We formalize this risk as a fundamental limit of efficient inference.

\begin{theorem}[The Over-Condensation Limit]
Let $q: \mathcal{M} \to \mathcal{M}/\sim$ be the condensation map. If the resolution of the condensed manifold $\delta(\mathcal{M}')$ falls below the causal resolution of the environment $\delta(\mathcal{W})$, such that distinct causal chains $c_1, c_2 \in \mathcal{W}$ map to the same vertex $v \in \mathcal{M}'$, then the probability of hallucination $P(H)$ is non-zero and irreducible.
\begin{equation}
P(H) = 1 - \frac{|\mathcal{H}_{\text{even}}(\mathcal{M}')|}{|\mathcal{H}_{\text{even}}(\mathcal{W})|}
\end{equation}
\end{theorem}
The above theorem states that hallucination is not a bug of randomness, but a structural consequence of lossy compression \cite{ma2022principles}. As the brain optimizes for speed (minimizing $\beta_1$ loops), it aggressively merges $\beta_0$ components. Any intelligence system, artificial or biological, might create a permanent geometric error when mergeing two components that reality keeps separate.

\noindent\textbf{The Certainty Paradox}
Why are hallucinations often delivered with high confidence?
In standard probabilistic models, uncertainty implies high variance. In MAI, uncertainty is path length.
The state of uncertainty (``I don't know'') is characterized by long search path, high energy, and failure to close cycle.
Because a hallucination is the result of a collapsed metric (a wormhole), the navigational cost to reach the false conclusion is near zero.
Physiologically, the system experiences the low metabolic cost of the hallucinatory path exactly as it experiences a valid memory. To the condensed brain, a hallucination \cite{allen2008hallucinating} is indistinguishable from a fact, because both define a geodesic of minimal action. The geometry of certainty does not measure truth but measures topological smoothness.

\section{Discussion and Conclusion} \label{sec:discussion}

We have proposed a new theoretical framework for cognitive computation: MAI, based on the homological parity principle and recursive topological condensation. By treating intelligence as a recursive topological operation, transforming the high-entropy flow of search into the low-entropy scaffold of memory, we provide a geometric explanation for the brain's ability to transcend the thermodynamic limits of ergodic search.

\noindent\textbf{MAI vs. Transformers: Hard Topology vs. Soft Attention} 
Current state-of-the-art AI, specifically the Transformer architecture \cite{vaswani2017attention}, relies on the \emph{Attention Mechanism}: a soft, $O(N^2)$ weighting of all pairwise relationships in a context window. While powerful, attention is topologically distinct from condensation. 1) Attention (Soft): The system maintains the full context window active in working memory. It attends to relevant tokens but does not structurally eliminate the irrelevant ones. The computational cost scales with context length \cite{radford2019language}. 2) Condensation (Hard): MAI performs a hard topological surgery. By condensing a sequence into a vertex, it physically removes the path from the computation. The cost becomes O(1), independent of context length. We argue that transformers effectively simulate the Search/Flow phase ($\Hodd$) but lack a true condensation operator. They are eternal novices, forced to reattend to the full context for every inference. A true Artificial General Intelligence (AGI) \cite{bubeck2023sparks} must implement recursive condensation to move beyond quadratic attention costs and achieve the linear-to-exponential scaling of the biological cortex.

\noindent\textbf{Physical Realization: The Topological Computer} 
The distinction between Software and Hardware vanishes at the limit of efficiency. Our framework suggests that the next generation of neuromorphic hardware should resemble topological insulators more closely than silicon logic gates \cite{hasan2010colloquium}.
In these materials, the Bulk-Boundary correspondence \cite{trifunovic2019higher} mirrors the Scaffold-Flow duality: 1) The Bulk (Insulating Scaffold): Represents the stable, condensed memory ($\Heven$), protected from energetic dissipation; 2) The Edge (Conducting Flow): Represents the active inference channel ($\Hodd$), where signals travel via protected states without backscattering. Implementing MAI on topologically protected substrates (e.g., Majorana zero modes \cite{sarma2015majorana}) would realize the ultimate thermodynamic efficiency: dissipationless inference, where the retrieval of memory costs zero thermal energy, leaving the energy budget entirely for the creation of new knowledge (search).

\noindent\textbf{Conclusion: The Geometry of Certainty}
Intelligence is the art of trading time for space. Faced with the time barrier of exponential search, the biological brain does not fight with raw speed; it sidesteps the battle by building a Tower of Scaffolds. Through the topological trinity (Search $\to$ Closure $\to$ Condensation), the cortex recursively reifies dynamic processes into static objects. The Geometry-of-Certainty framework allows finite agents to model infinite worlds. However, the efficiency comes with an irreducible limit. The wormholes that allow us to leap instantly from premise to conclusion are the same geometric defects that lead to hallucination. When we over-condense the manifold, we lose the resolution to distinguish truth from plausible fiction. Ultimately, the limit of an intelligent system is not its processing speed, but its topological resolution, the fineness of the scaffold against the entropy of the flow. To evolve is to build a higher scaffold; to learn is to climb it.

\bibliography{reference}

\bibliographystyle{IEEEtran}

\end{document}